\documentclass[conference]{IEEEtran}
\IEEEoverridecommandlockouts
\usepackage{cite}
\usepackage{amsmath,amssymb,amsfonts}
\usepackage{algorithmic}
\usepackage{graphicx}
\usepackage{textcomp}
\usepackage{xcolor}
\def\BibTeX{{\rm B\kern-.05em{\sc i\kern-.025em b}\kern-.08em
    T\kern-.1667em\lower.7ex\hbox{E}\kern-.125emX}}
\begin{document}

\title{Digital Controllers in Discrete and Continuous Time Domains for a Robot Arm Manipulator \\
}

\author{\IEEEauthorblockN{1\textsuperscript{st} Dhiman Chowdhury, \textit{Student Member, IEEE}}
\IEEEauthorblockA{\textit{Electrical Engineering} \\
\textit{University of South Carolina}\\
Columbia, South Carolina, USA\\
dhiman@email.sc.edu}
\and
\IEEEauthorblockN{2\textsuperscript{nd} Mrinmoy Sarkar, \textit{Student Member, IEEE}}
\IEEEauthorblockA{\textit{Electrical and Computer Engineering} \\
\textit{North Carolina A \& T State University}\\
Greensboro, NC 27411, USA \\
msarkar@aggies.ncat.edu}}
%\and
%\and
%\and
%\and

\maketitle

\begin{abstract}
This paper articulates design and performance analysis of digital controllers in discrete and continuous time domains for a single-joint robot arm manipulator. The investigated robot arm system is modeled as a single degree of freedom (DOF) plant and there is a feedback sensor implying a closed-loop system. The design approach incorporates discrete (z-plane) and continuous time (warped s-plane or w-plane) domain parameters. Four digital controllers - phase-lag, phase-lead, proportional-integral (PI) and proportional-integral-derivative (PID) are theoretically designed and implemented to achieve a phase margin of 40 deg. for the compensated system. For performance evaluations, Bode plots of the compensated open-loop systems and step response characteristics of the closed-loop systems are determined.
\end{abstract}


\section{Introduction}
Controllers are essential to determine the changes of system parameters and to attain desired characteristics with performance specifications, which are related to steady-state accuracy, transient response, stability and disturbance reduction \cite{1}. Analog controllers are hard to synthesize complicated logics, to make dynamic interfaces among multiple subsystems and are highly susceptible to corruption by extraneous noise sources \cite{1}. However, digital controllers are reliable, since no signal loss occurs in analog-to-digital (A/D) and digital-to-analog (D/A) conversions and are more flexible and accurate in case of sophisticated logic implementation \cite{1}. In addition, digital controllers are not subject to external noises. Several applications of digital control algorithms in robotics and automated systems are reported in \cite{1} - \cite{5}.

This paper documents design methodologies of different types of digital controllers - phase-lag, phase-lead, proportional-integral (PI) and proportional-integral-derivative (PID) for a physical system of a single DOF robot arm manipulator. Phase margin is compensated in the system employing cross-over frequency as the primary design parameter. Bode diagrams of the open-loop systems and step response characteristics of the closed-loop systems are determined. Comparative analysis of the designed controllers is carried out. MATLAB \textregistered\; simulations are applied for the proposed framework. However, several works on robot arm systems for efficient control phenomena are articulated in \cite{6} - \cite{10}.

The major contributions of this research paper are to - \\
a. conceptualize digital controllers in discrete and continuous time domains for a single-joint robot arm manipulator. \\
b. design controllers for a particular compensation criterion (phase margin) using frequency response techniques.\\
c. derive Bode plots and step responses of the compensated systems and compare the characteristics obtained from different controllers.

The remainder of this manuscript is organized as follows. Section II documents the investigated robot arm system, plant transfer function and basic compensation theory of the digital controllers. Sections III - VI describe the design methodologies of phase-lag, phase-lead, PI and PID controllers, respectively. Section VII presents the step response characteristics. Section VIII concludes the paper.

\begin{figure}[!b]
	\centering
	\includegraphics[height=1.3in ,width=3.5 in]{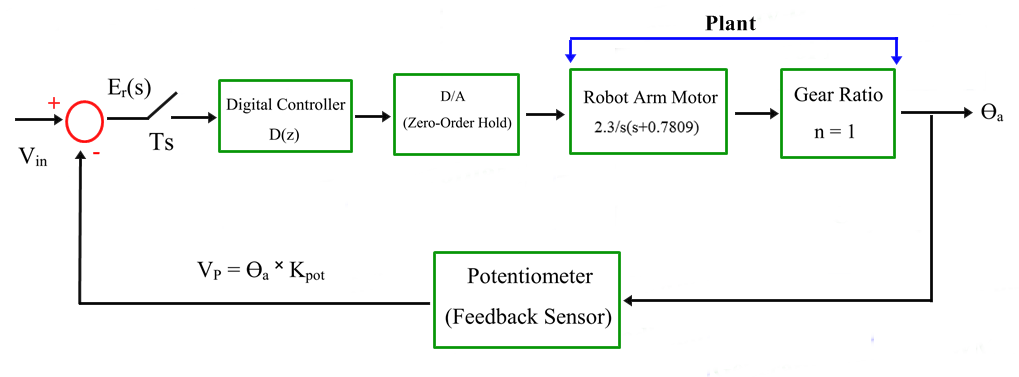}
	\caption{Block diagram of the robot arm manipulator system}
	\label{fig1}
\end{figure}

\section{Robot Arm System, Plant Transfer Function and Digital Controller}
\begin{figure}[!t]
	\centering
	\includegraphics[height=2in ,width=3 in]{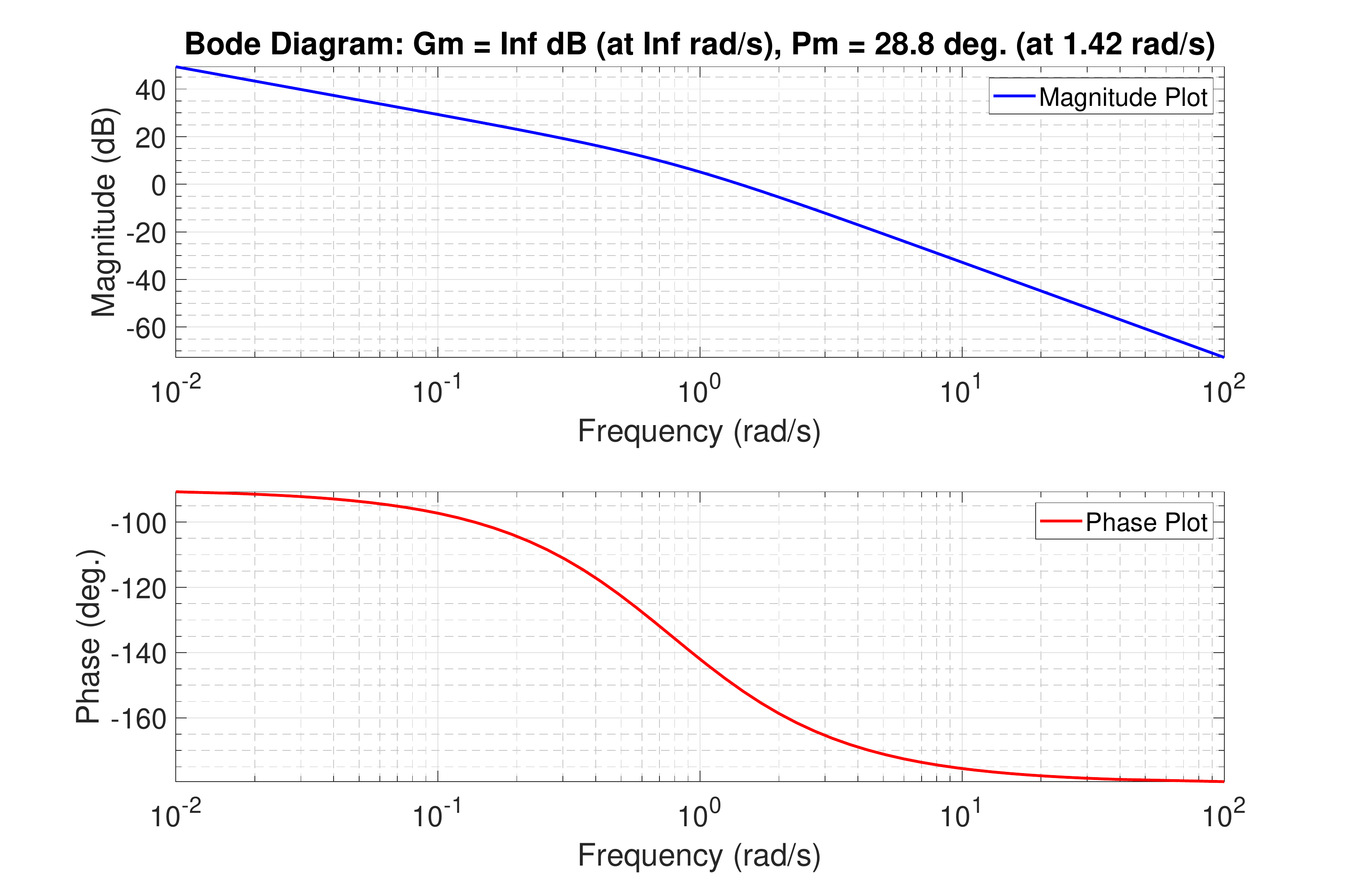}
	\caption{Bode plot of the uncompensated plant (robot arm motor + gear)}
	\label{fig2}
\end{figure}

\begin{figure}[!t]
	\centering
	\includegraphics[height=2in ,width=3 in]{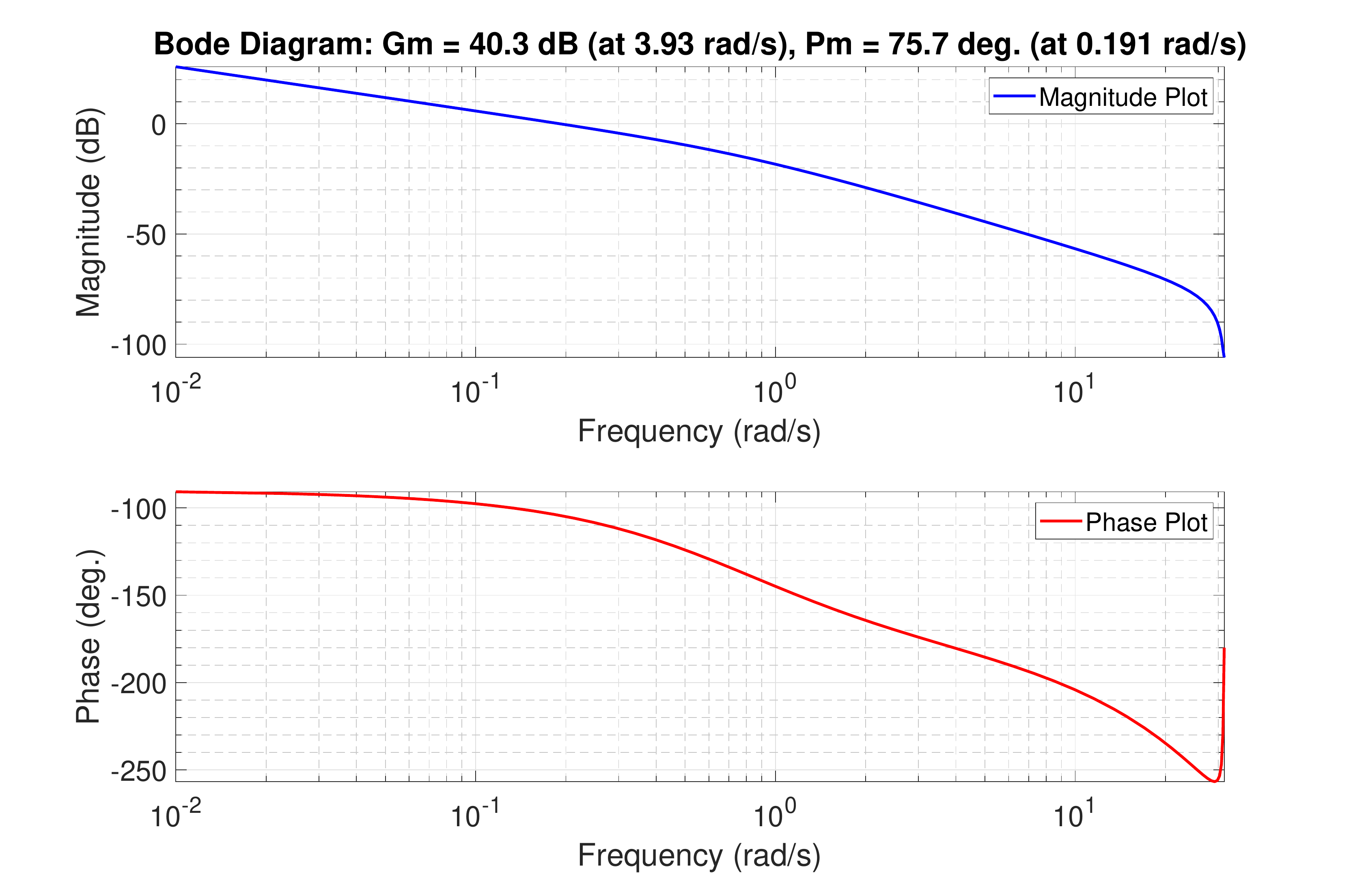}
	\caption{Bode plot of the plant + feedback sensor (uncompensated open-loop system)}
	\label{fig3}
\end{figure}

Fig. \ref{fig1} presents the block diagram of the investigated single joint (1-DOF) robot arm manipulator system. The plant configuration follows the mathematical modeling approach reported in \cite{6} - \cite{7}. Here the sampling time is  $Ts=0.1$ s and the potentiometer constant $K_{pot} =\frac{V_{p}}{\theta_{a}} = \frac{12-0}{180-0}=0.0667$ V/deg. However, the compensation theory, corresponding mathematical derivations for the controller design and open loop and closed loop parameters of the controllers described in this paper follow the literature documented in \cite{1}. According to \cite{1}, for a first-order compensation, the controller transfer function can be expressed as $D(z)=\frac{K_{d}(z-z_{0})}{z-z_{p}}$. Here $z_{0}$ and $z_{p}$ are the respective zero and pole locations. The bilinear or trapezoidal transformation of the controller from the discrete z-plane to the continuous w-plane (warped s-plane) implies $D(w)=D(z), z=\frac{1+(T/2)w}{1-(T/2)w}$, such that $D(w)=a_{0}\frac{1+(w/\omega_{w0})}{1+(w/\omega_{wp})}$. 
Here $\omega_{w0}$ and $\omega_{wp}$ are the respective zero and pole locations in the w-plane and $a_{0}$ is the compensator DC gain. According to the bilinear approximation, $w=\frac{2}{T}\frac{z-1}{z+1}$. From the above equations, in z-plane the controller can be realized as
\begin{equation}
D(z)=a_{0}\frac{\omega_{wp}(\omega_{w0}+2/T)}{\omega_{w0}(\omega_{wp}+2/T)}\frac{z-(\frac{2/T-\omega_{w0}}{2/T+\omega_{w0}})}{z-(\frac{2/T-\omega_{wp}}{2/T+\omega_{wp}})}
\end{equation}
For the uncompensated plant, the controller, $D(z)=1$. The zero-order hold transfer function can be defined as $G_{HO}(s)=\frac{1-e^{-sT}}{s}$. The continuous-time plant with feedback sensor gain transfer function, $G_{c}(s)=\frac{0.1533}{s^2+0.7809s}$. 
The discrete-time plant with feedback sensor gain transfer function is
\begin{equation}
G_{d}(z)=\frac{0.0007471z+0.0007279}{z^2-1.925z+0.9249}
\end{equation}
Figs. \ref{fig2} and \ref{fig3} present the Bode plots of the plant and uncompensated plant + feedback sensor system, respectively.

\section{Phase-Lag Controller Design}
\begin{figure}[!t]
	\centering
	\includegraphics[height=2in ,width=3 in]{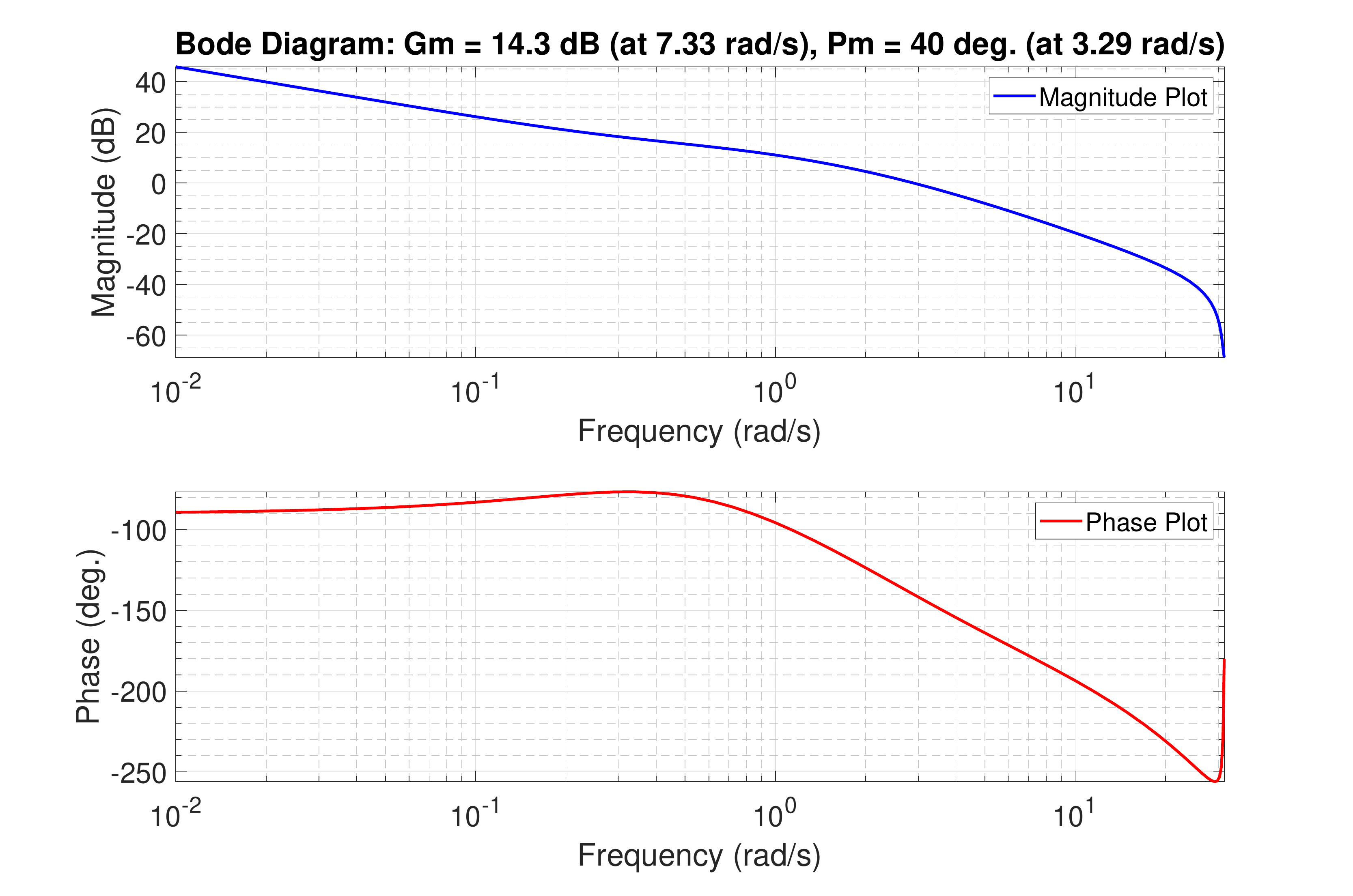}
	\caption{Bode plot of the compensated open-loop system using phase-lag controller}
	\label{fig5}
\end{figure}

The DC gain of the lag controller design, $a_{0}=10$ and the high-frequency gain can be expressed as $G_{hf}(dB)=20log\frac{a_{0}\omega_{wp}}{\omega_{w0}}$. The maximum phase shift lies between 0 and -90 deg. which depends on the ratio $\omega_{w0}/\omega_{wp}$. In this paper, the controller is designed for 40 deg. phase margin and the cross-over or phase margin frequency for this design is selected as $\omega_{wc}=3.29$  $rads^{-1}$. Here $\omega_{w0}=0.1\omega_{wc}$, and $\omega_{wp}=\frac{\omega_{w0}}{a_{0}|G_{d}(j\omega_{wc})|}$. 
The design approximates that the controller introduces 5 deg. phase lag to the system and $|D(j\omega_{wc})G_{d}(j\omega_{wc})|=1$. The lag controller implies that $\omega_{w0}=0.3290<\omega_{wp}=2.3979$ and the compensating phase angle, $\phi_{m}=(-180+5+40)=-135$ deg. The controller transfer function is $D_{lag}(z)=\frac{66.15z-64.01}{z-0.7859}$.
Fig. \ref{fig5} presents the Bode plot of the lag-compensated open loop system. It can be observed that the phase margin of the compensated plant is $P_{m}=40$ deg. at 3.29 $rads^{-1}$ and the gain margin is $G_{m}=14.3$ dB.
The phase-lag controller reduces the phase margin by $(75.7-40)=35.7$ deg.
%\begin{figure}[!t]
%	\centering
%	\includegraphics[height=2in ,width=3 in]{bode_only_lag}
%	\caption{Bode plot of the phase-lag controller}
%	\label{fig4}
%\end{figure}

\section{Phase-Lead Controller Design}
\begin{figure}[!t]
	\centering
	\includegraphics[height=2in ,width=3 in]{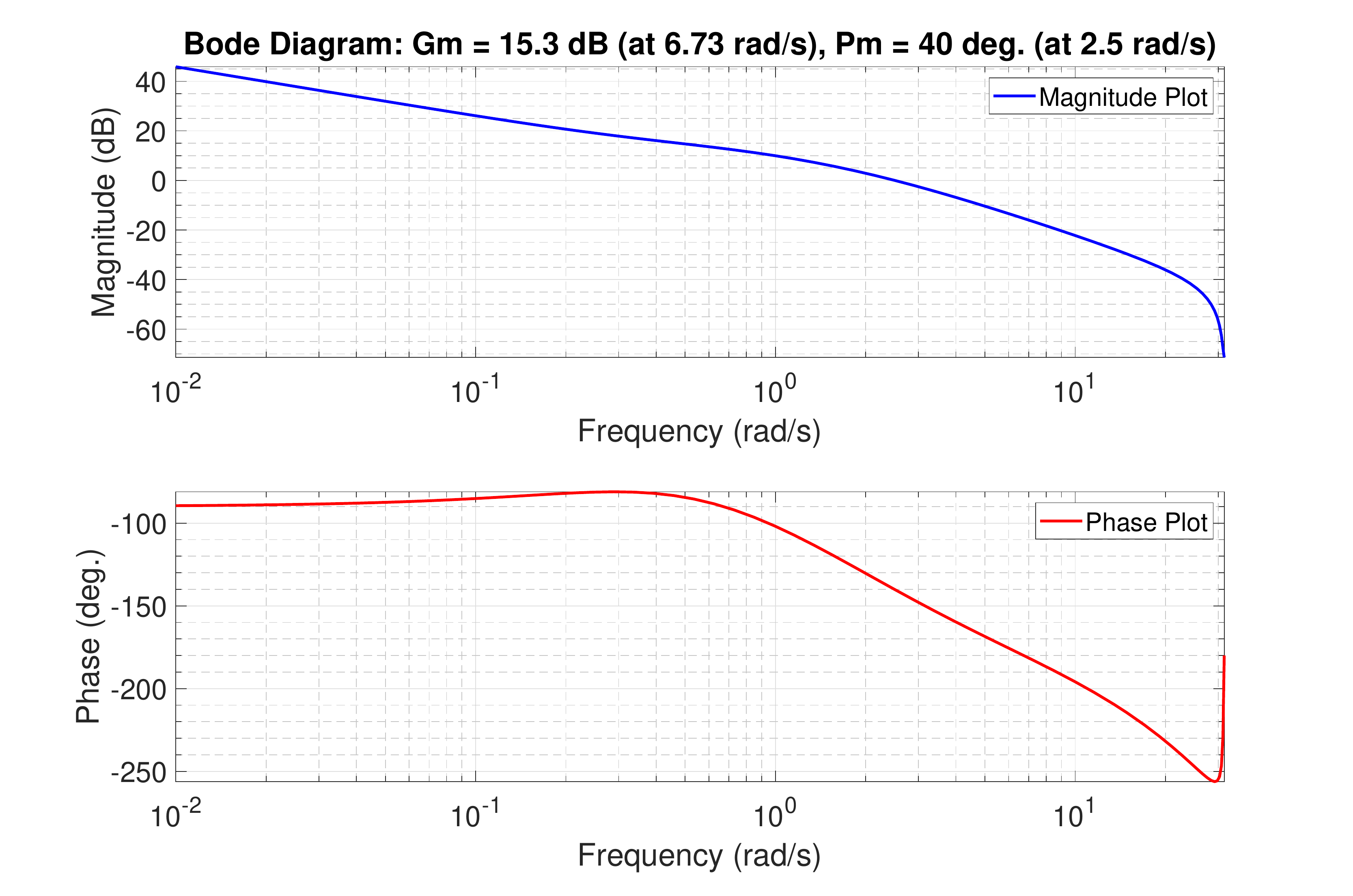}
	\caption{Bode plot of the compensated open-loop system using phase-lead controller}
	\label{fig7}
\end{figure}

%\begin{figure}[!t]
%	\centering
%	\includegraphics[height=2in ,width=3 in]{bode_only_lead}
%	\caption{Bode plot of the phase-lead controller}
%	\label{fig6}
%\end{figure}
Again, the DC gain of the phase-lead controller, $a_{0}=10$ and in this paper, the controller is designed for 40 deg. phase margin and the cross-over or phase margin frequency for this design is selected as $\omega_{wc}=2.5$  $rads^{-1}$. The lead controller design approach yields to
\begin{equation}
D(j\omega_{wc})G_{d}(j\omega_{wc})=1\angle(180+\phi_{pm}), |D(j\omega_{wc})|=\frac{1}{|G_{d}(j\omega_{wc})|}
\end{equation}
here $\phi_{pm}$ is the desired phase margin and
\begin{equation}
D(w)=a_{0}\frac{1+w/(a_{0}/a_{1})}{1+w/(b_{1})^-1}
\end{equation}
where $\omega_{w0}=\frac{a_{0}}{a_{1}}$ and $\omega_{wp}=\frac{1}{b_{1}}$. The angle associated with the controller can be expressed as
\begin{equation}
\theta_{r}=\angle D(j\omega_{wc})=180+\phi_{pm}-\angle G_{d}(j\omega_{wc})
\end{equation}
According to \cite{1}, it can be evaluated that
\begin{equation}
a_{1}=\frac{1-a_{0}|G_{d}(j\omega_{wc})|\cos\theta_{r}}{\omega_{wc}|G_{d}(j\omega_{wc})|\sin\theta_{r}}, b_{1}=\frac{\cos\theta_{r}-a_{0}|G_{d}(j\omega_{wc})|}{\omega_{wc}\sin\theta_{r}}
\end{equation}
In the design procedure, $\omega_{wc}$ is selected to satisfy the constraints: $\angle G_{d}(j\omega_{wc})<180+\phi_{pm}; |D(j\omega_{wc})|>a_{0}$, $|G_{d}(j\omega_{wc})|<\frac{1}{a_{0}}; b_{1}>0$ and $\cos\theta_{r}>a_{0}|G_{d}(j\omega_{wc})|$. The lead controller implies that $\omega_{w0}=0.3641 <\omega_{wp}=1.9603$. The calculated design parameters: $a_{1}=27.4649$, $b_{1}=0.5101$, $\theta_{r}=389.8149$ deg. and $\cos\theta_{r}=0.9671$.
The controller transfer function is $ D_{lead}(z)=\frac{49.927(z-0.9642)}{(z-0.8251)}$.
%\begin{equation}
%D_{lead}(z)=\frac{15.809(z-0.8771)}{(z-0.8057)}
%\end{equation}
Fig. \ref{fig7} presents the Bode plot of the lead-compensated open loop system. It can be implied that the phase margin of the compensated plant is $P_{m}=40$ deg. at 2.5 $rads^{-1}$ and the gain margin is $G_{m}=15.3 $ dB. The phase-lead controller reduces the phase margin by $(75.7-40)=35.7$ deg.
\section{Proportional-Integral (PI) Controller Design}
\begin{figure}[!t]
	\centering
	\includegraphics[height=2in ,width=3 in]{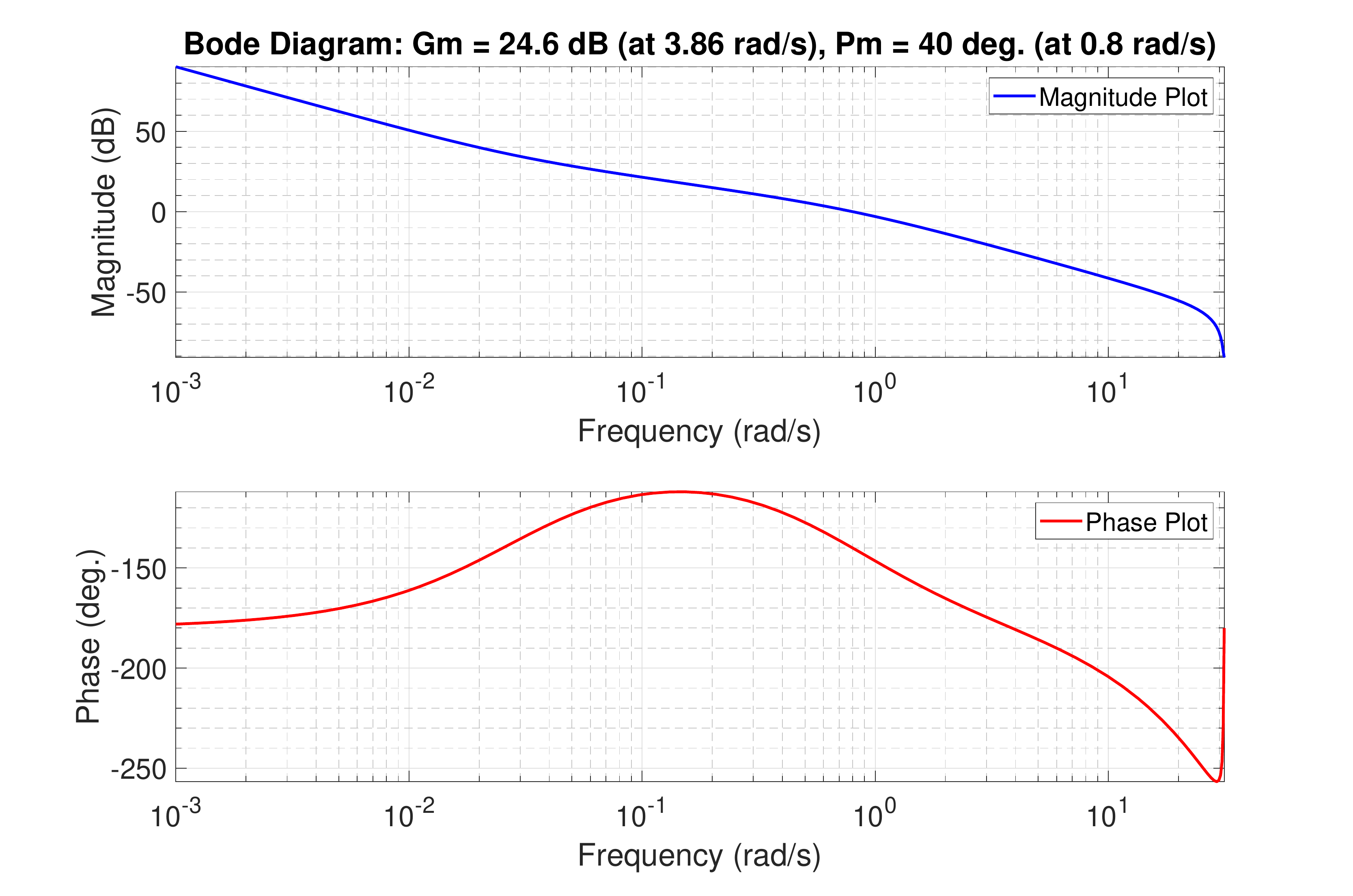}
	\caption{Bode plot of the compensated open-loop system using proportional-integral (PI) controller}
	\label{fig9}
\end{figure}

The PI controller transfer function can be expressed as $D(w)=K_{P}+\frac{K_{I}}{w}=K_{I}\frac{1+w/\omega_{w0}}{w}$,
here $\omega_{w0}=K_{I}/K_{P}$. However, the discrete transfer function of a PI controller can be expressed as $D(z)=K_{P}+K_{I}\frac{T}{2}\frac{z+1}{z-1}$, and the controller frequency response is
\begin{equation}
D(j\omega_{w})=K_{P}-j\frac{K_{I}}{\omega_{w}}=|D(j\omega_{w})|e^{j\theta_{r}}
\end{equation}
At the cross-over frequency, the controller yields to
\begin{equation}
D(j\omega_{wc})G_{d}(j\omega_{wc})=1\angle(-180+\phi_{pm})
\end{equation}
At the cross-over frequency 0.8 $rads^{-1}$ in this work,
\begin{equation}
K_{P}-j\frac{K_{I}}{\omega_{wc}}=|D(j\omega_{wc})|(\cos\theta_{r}+j\sin\theta_{r})
\end{equation}
However, the phase angle associated with the controller is
\begin{equation}
\theta_{r}=-180+\phi_{pm}-\angle G_{d}(j\omega_{wc})
\end{equation}
The coefficients can be expressed as
\begin{equation}
K_{P}=\frac{\cos\theta_{r}}{|G_{d}(j\omega_{wc}|}, K_{I}=-\frac{\omega_{wc}\sin\theta_{r}}{|G_{d}(j\omega_{wc})|}
\end{equation}
The controller transfer function is calculated as
\begin{equation}
D_{PI}(z)=\frac{5.839z -5.823}{z-1}
\end{equation}
The design parameters are: $\theta_{r}=357.9840$ deg., $K_{P}=5.8307$ and $K_{I}=0.1642$. Fig. \ref{fig9} presents the Bode plot of the PI-compensated open loop system. The phase margin of the compensated plant is $P_{m}=40$ deg. at 0.8 $rads^{-1}$ and the gain margin is $G_{m}=24.6$ dB. The PI controller reduces the phase margin by $(75.7-40)=35.7$ deg.

%\begin{figure}[!t]
%	\centering
%	\includegraphics[height=2in ,width=3 in]{bode_only_PI}
%	\caption{Bode plot of the proportional-integral (PI) controller}
%	\label{fig8}
%\end{figure}

\begin{figure}[!b]
	\centering
	\includegraphics[height=2in ,width=3 in]{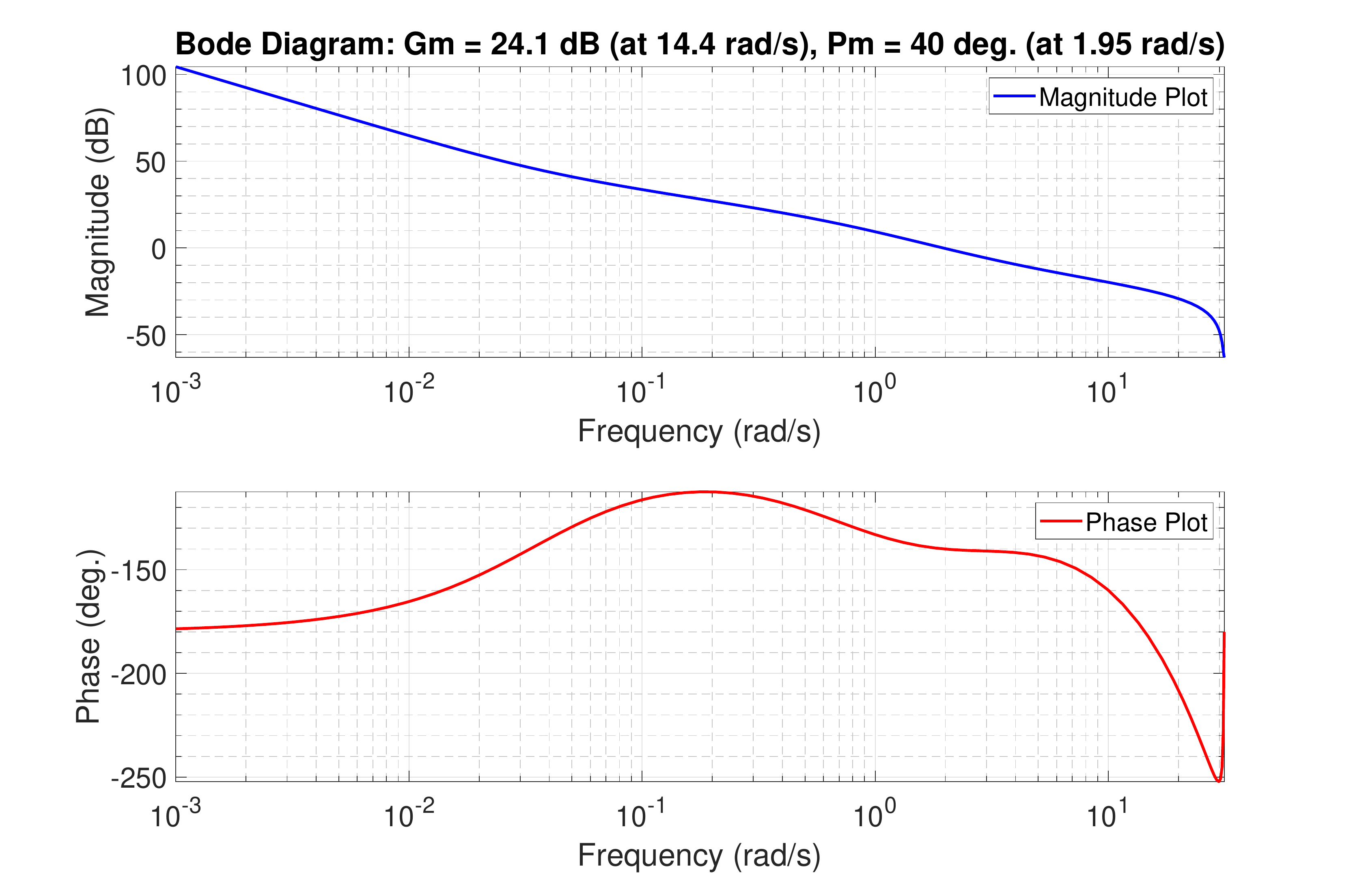}
	\caption{Bode plot of the compensated open-loop system using proportional-integral-derivative (PID) controller}
	\label{fig11}
\end{figure}
\section{Proportional-Integral-Derivative (PID) Controller Design}
%\begin{figure}[!t]
%	\centering
%	\includegraphics[height=2in ,width=3 in]{bode_only_PID}
%	\caption{Bode plot of the proportional-integral-derivative (PID) controller}
%	\label{fig10}
%\end{figure}

The PID controller transfer function can be expressed as $D(w)=K_{P}+\frac{K_{I}}{w}+K_{D}w$.
The discrete transfer function of a PID controller can be expressed as $D(z)=K_{P}+K_{I}\frac{T}{2}\frac{z+1}{z-1}+K_{D}\frac{z-1}{Tz}$.
The controller frequency response is
\begin{equation}
D(j\omega_{w})=K_{P}+j(K_{D}\omega_{w}-\frac{K_{I}}{\omega_{w}})=|D(j\omega_{w})|e^{j\theta_{r}}
\end{equation}
At the cross-over frequency 1.95 $rads^{-1}$ in this design procedure,
\begin{equation}
K_{P}+j(K_{D}\omega_{wc}-\frac{K_{I}}{\omega_{wc}})=|D(j\omega_{wc})|(\cos\theta_{r}+j\sin\theta_{r})
\end{equation}
From the above equations, it can be derived as
\begin{equation}
K_{P}=\frac{\cos\theta_{r}}{|G_{d}(j\omega_{wc}|}, K_{D}\omega_{wc}-\frac{K_{I}}{\omega_{wc}}=\frac{\sin\theta_{r}}{|G_{d}(j\omega_{wc})|}
\end{equation}
As a design consideration, adding a pole in the derivative term modifies the controller transfer function as
\begin{equation}
D(w)=K_{P}+\frac{K_{I}}{w}+\frac{K_{D}w}{1+(T/2)w}
\end{equation}
However, the modified frequency response is
\begin{equation}
D(j\omega_{w})=K_{P}-j\frac{K_{I}}{\omega_{w}}+\frac{K_{D}j\omega_{w}}{1+j\omega_{w}(T/2)}
\end{equation}
which yields to
\begin{equation}
[K_{P}+\frac{K_{D}\omega_{wc}^2(2/T)}{(2/T)^2+\omega_{wc}^2}]+j[\frac{K_{D}\omega_{wc}(2/T)^2}{(2/T)^2+\omega_{wc}^2}-\frac{K_{I}}{\omega_{wc}}] = K_{R} + jK_{C}
\end{equation}
Here $K_{R}=\frac{\cos\theta_{r}}{|G_{d}(j\omega_{wc})|}$ and $K_{C}=\frac{\sin\theta_{r}}{|G_{d}(j\omega_{wc})|}$.
Thereby, it can be concluded that
\begin{equation}
K_{P}+\frac{K_{D}\omega_{wc}^2(2/T)}{(2/T)^2+\omega_{wc}^2}=\frac{\cos\theta_{r}}{|G_{d}(j\omega_{wc})|}
\end{equation}
and
\begin{equation}
\frac{K_{D}\omega_{wc}(2/T)^2}{(2/T)^2+\omega_{wc}^2}-\frac{K_{I}}{\omega_{wc}}=\frac{\sin\theta_{r}}{|G_{d}(j\omega_{wc})|}
\end{equation}
The controller transfer function is calculated as
\begin{equation}
D_{PID}(z)=\frac{81.51z^2-139.5z+58.07}{z^2-z}
\end{equation}

The controller parameters are $K_{I}=0.85$, $K_{D}=5.8069$ and $K_{P}=23.3942$, respectively. The angle associated with the controller is $\theta_{r}=383.7620$ deg. Fig. \ref{fig11} presents the Bode plot of the PID-compensated open loop system. The phase margin of the compensated plant is $P_{m}=40$ deg. at 1.95 $rads^{-1}$ and the gain margin is $G_{m}=24.1$ dB. The PID controller reduces the phase margin by $(75.7-40)=35.7$ deg.

\section{Step Response Characteristics}
\begin{figure}[!t]
	\centering
	\includegraphics[height=2in ,width=3 in]{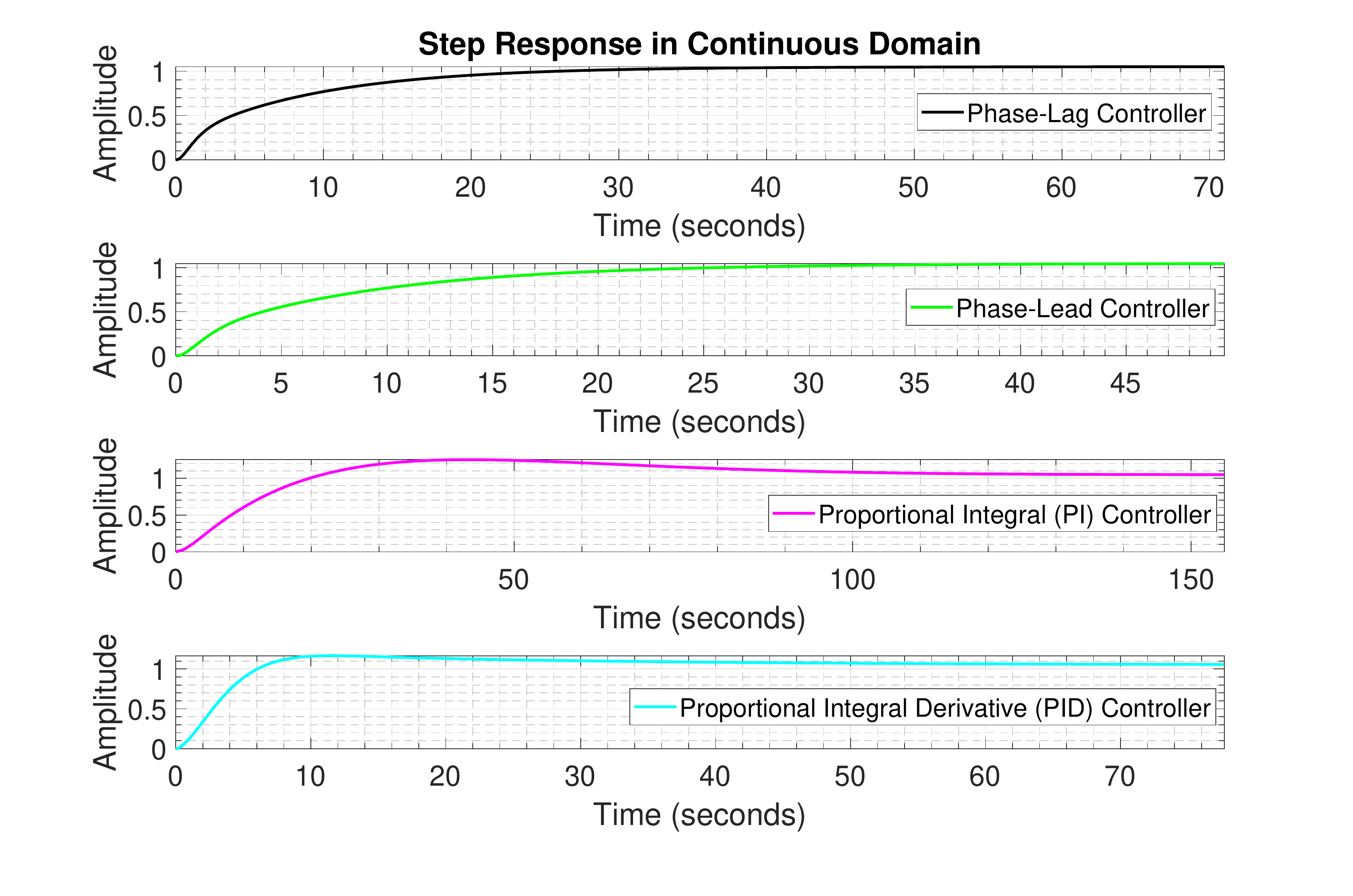}
	\caption{Step response of the compensated closed-loop system in continuous time domain using different controllers}
	\label{fig12}
\end{figure}

The single-joint robot arm manipulator in this paper has an input of $V_{in}=0.07u(t)$; where $u(t)$ is the unit step function. Fig. \ref{fig12} presents the scaled step response of the compensated closed loop system in continuous time domain for the digital controllers. However, Table \ref{table-1} shows the step response characteristics of the controllers for comparative analysis. The steady-state error for all the controllers is found as zero. The phase-lag and phase-lead controllers produce zero percent overshoot. However, the rise time for the PID controller is the lowest among all, whereas the phase-lead controller produces the lowest settling time.

\begin{table}[!t]
	\centering
	\caption{Step Response Characteristics of the Controllers}
	\begin{tabular}{|p{1.1in}||p{0.35in}||p{0.35in}||p{0.35in}||p{0.35in}|}
		\hline
		Characteristics & Lag & Lead & PI & PID \\
		\hline
		Steady-State Error & 0 & 0 & 0 & 0 \\
		\hline
		Percent Overshoot (\%) & 0 & 0 & 19.1 & 11.6 \\
		\hline
		Rise Time (s) & 18.4 & 17.9 & 15.5 & 4.6 \\
		\hline
		Settling Time (s) & 34.1 & 33 & 106 & 51.9 \\ \hline
	\end{tabular}%
	\label{table-1}%
\end{table}%

\section{Conclusion}
This paper presents design and performance analysis of digital controllers - phase-lag, phase-lead, PI and PID for a single DOF robot arm manipulator. The phase margin for the compensated system is selected as 40 deg and cross-over frequency is the prime criterion in the design procedure. However, the controllers are designed and implemented in both discrete (z-domain or actual digital) and continuous (warped s-domain or w-plane) time domains. For performance assessments, MATLAB \textregistered\; simulations are carried out for open-loop Bode plots and closed loop step response characteristics.

\end{document}